\documentclass[letterpaper]{article} 
\usepackage{aaai2026}  
\usepackage{times}  
\usepackage{helvet}  
\usepackage{courier}  
\usepackage[hyphens]{url}  
\usepackage{graphicx} 
\usepackage{natbib}  
\usepackage{caption} 
\usepackage{algorithm}
\usepackage{algorithmic}

\usepackage{newfloat}
\usepackage{listings}
\DeclareCaptionStyle{ruled}{labelfont=normalfont,labelsep=colon,strut=off} 
\floatstyle{ruled}
\newfloat{listing}{tb}{lst}{}
\floatname{listing}{Listing}
\usepackage{graphicx}
\usepackage{multirow}
\usepackage{amsmath,amssymb,amsfonts,amsthm}
\usepackage{caption}
\usepackage{subcaption}
\usepackage{cleveref}

\crefformat{section}{\S#2#1#3}
\crefformat{subsection}{\S#2#1#3}
\crefformat{subsubsection}{\S#2#1#3}
\usepackage{ctable}

\newcommand{\todo}[1]{{\color{red}\textbf{[TODO: #1]}}}

\newcommand{\paper}{paper }

\title{LLA: Enhancing Security and Privacy for\\
Generative Models with \underline{L}ogic-\underline{L}ocked \underline{A}ccelerators}

\author{
    You Li\equalcontrib, Guannan Zhao\equalcontrib, Yuhao Ju, Yunqi He, Jie Gu, Hai Zhou\\
}
\affiliations{
    Northwestern University, Evanston, IL, USA

    \{you.li, gnzhao, yuhaoju2017, yunqi.he\}@u.northwestern.edu,
    \{jgu, haizhou\}@northwestern.edu
}

\usepackage{bibentry}

\begin{document}
\setlength{\abovecaptionskip}{7pt} 

\maketitle

\begin{abstract}
We introduce LLA, an effective intellectual property (IP) protection scheme for generative AI models. LLA leverages the synergy between hardware and software to defend against various supply chain threats, including model theft, model corruption, and information leakage. On the software side, it embeds key bits into neurons that can trigger outliers to degrade performance and applies invariance transformations to obscure the key values. On the hardware side, it integrates a lightweight locking module into the AI accelerator while maintaining compatibility with various dataflow patterns and toolchains. An accelerator with a pre-stored secret key acts as a license to access the model services provided by the IP owner. 
The evaluation results show that LLA can withstand a broad range of oracle-guided key optimization attacks, while incurring a minimal computational overhead of less than 0.1\% for 7,168 key bits.
\end{abstract}

\section{Introduction}

Generative AI (GenAI) is a revolutionary technology that automatically creates a variety of content, including text, images, videos, and audio, in response to users' input. Training generative models requires massive data, substantial training resources, and specialized expertise. As a result,   new business models have emerged to allow clients with limited resources to access GenAI capabilities. For instance, Inference-as-a-Service (\textit{IaaS}) refers to a deployment approach in which generative models are hosted on cloud platforms and accessed through APIs. In contrast, \textit{self-hosting} enables organizations to deploy models internally, offering greater control over confidential information. 
In such situations, model parameters are directly exposed to various participants in the \textit{supply chain}, including cloud service providers, network operators, and end-users.
Unauthorized use or distribution of proprietary models can cause significant financial losses to intellectual property (IP) owners.
Moreover, unrestricted access to model parameters poses serious security risks.
For example, adversarial prompts can cause the generation of harmful or misleading content~\cite{zhuang2023pilot,zou2023universal}, while backdoor attacks can manipulate generated outputs through malicious triggers embedded in compromised models~\cite{zhao2023prompt,chou2023backdoor}.

\textit{Model locking} leverages the principle of \textit{hardware root-of-trust} to defend against model theft and various supply chain threats.
It inserts key-controlled protection units into the neural network, with the secret key stored on hardware in a tamper-proof module~\cite{nicholas2021secure,chakraborty2009harpoon,kamali2018lut}. Without knowing the correct key, an adversary cannot restore the original functionality of the model.
The main advantages of model locking are three-fold. Firstly, it is a proactive approach that offers guaranteed security. In contrast, reactive approaches such as watermarking~\cite{kirchenbauer2023watermark} and fingerprinting~\cite{xu2024instructional} can provide evidence of model ownership but cannot prevent unauthorized use of the model. Secondly, model locking introduces minimal computational and hardware overhead. Other proactive approaches, such as TEE-based execution~\cite{mo2020darknetz}, parameter encryption~\cite{lin2020chaotic}, and memory encryption~\cite{zuo2021sealing}, provide high levels of security but come with significant computational or hardware costs. Lastly, model locking assumes a general and realistic threat model, allowing the architecture and all model parameters to be publicly released. This allows the IP owner to host the model on a cloud platform or send it to an end-user without compromising security or ownership.

This \paper presents LLA, a comprehensive model locking framework for generative models (Fig.~\ref{fig:flow_1}). 
LLA achieves effectiveness, robustness, and efficiency simultaneously through an integration of software and hardware components.
In the software domain, LLA identifies feature outliers within an FFN module and inserts key bits to manipulate these outliers. As such, it can cause substantial degradation in model performance with a small number of key bits.
Moreover, it applies invariant transformations to obfuscate key values, thus thwarting a wide range of oracle-guided attacks at minimal cost.
In the hardware domain, LLA embeds a lightweight locking module within systolic array AI accelerators. This module is designed to be fully compatible with existing dataflow patterns and model formats.
A pre-activated AI accelerator can serve as a license to access all current and future services offered by the IP owner.

Our main contributions are as follows:

\noindent $\bullet$ We propose LLA, the first complete method to apply model locking on large generative models.

\noindent $\bullet$ We devise a systematic approach to
find and construct  
critical components within a generative model. Locking these components can significantly impair model functionality while preserving the confidentiality of key values.

\noindent $\bullet$ We develop a lightweight solution to enable the execution of LLA-protected models on general AI hardware.

\noindent $\bullet$ We conduct comprehensive experiments to evaluate the effectiveness of LLA across a diverse set of generative models, assess its efficiency on AI hardware, and examine its robustness against various attacks.

\section{Background and Related Work}
\label{sec:background}

\paragraph{IP Protection for AI Models.}
Researchers have proposed various methods to safeguard the IP of AI models. \textit{Watermarking} embeds hidden information into model parameters ~\cite{uchida2017embedding} or model outputs~\cite{adi2018turning} to prevent unauthorized use. Unfortunately, it can only passively verify the ownership of a model after it has been stolen or infringed. \textit{Model encryption}~\cite{zhou2023nnsplitter,mu2024encryip} prevents unauthorized access to model parameters with encryption and obfuscation techniques. However, an encrypted model must be restored to the original state before execution, which incurs high overhead and makes it vulnerable to side channel attacks~\cite{li2025licensenet}. \textit{Model splitting} isolates a subset of sensitive computations and executes them within a Trusted Execution Environment (TEE) of a processor~\cite{khan2021utilizing,wang2023building}. Its drawbacks include hardware cost, memory constraints, and performance overhead. \textit{Homomorphic encryption} can guarantee the security and privacy of AI models~\cite{gilad2016cryptonets,sun2018private}. However, due to the extremely high computational cost, it can barely be applied in practice.

Model locking originates from \textit{logic locking}~\cite{kamali2022advances}, which uses binary key bits to protect the IP of logic circuits.
Hardware-protected neural network (HPNN)~\cite{chakraborty2020hardware} performs model locking in the following steps: \textit{i)} selecting neurons at fixed locations of hidden layers as the \textit{protected neurons}; \textit{ii)} associating a key bit with each protected neuron to control whether to flip the pre-activation value; \textit{iii)} training the model as a function of a predetermined secret key. NN-Lock~\cite{alam2022nn,goldstein2021preventing} utilizes cryptographic primitives to obfuscate the model parameters. 
While these methods are highly efficient and robust, they have several limitations in common: \textit{i)} it is not clear how an encrypted model can be executed on general AI hardware; \textit{ii)} they are designed for discriminative models rather than large generative models. GenAI presents unique challenges to model locking, and standard model locking techniques are no longer effective as the model scales. For example, removing 25$\%$ of layers from Llama2-13B results in only a slight performance drop~\cite{men2024shortgpt}, and over 95$\%$ of neurons in FFN modules of OPT-175B are inactive during inference~\cite{lilazy}. 

\paragraph{Transformer Architecture.}
A decoder-only transformer is a stack of \textit{N} transformer \textit{blocks}, each consisting of two main modules: a multi-head self-attention module and a feed-forward network (FFN) module (Fig.~\ref{fig:flow_2}(a)). Let $\mathbf{X} \in \mathbb{R}^{T \times D_{\mathrm{m}}}$ denote the input matrix of both modules, where \textit{T} is the number of tokens and $D_{\mathrm{m}}$ is the number of features of each token. Following the self-attention module, the same FFN module is applied identically to each token. The FFN module comprises two linear layers (Fig.~\ref{fig:flow_2}(b)). The first layer, $\mathtt{up}$, expands
a token's feature vector from the model dimension $D_{\mathrm{m}}$ to the intermediate dimension $D_{\mathrm{ff}}$. Certain architectures incorporate a gated matrix $\mathbf{W}^{\mathtt{gate}}$ in parallel with the original up-projection matrix $\mathbf{W}^{\mathtt{up}}$, and the output vectors of the two matrices are subsequently combined through element-wise multiplication. A non-linear activation layer, $\mathtt{act}$, is applied between the two linear layers. Finally, the second layer, $\mathtt{down}$, projects the expanded vector from the intermediate space back to the \textit{hidden state space}.

\paragraph{Outliers in GenAI Models.}
Outliers refer to abnormally large values within GenAI models~\cite{dettmers2022gpt3,kovaleva2021bert}. \textit{Weight outliers} appear in specific columns of the down-projection matrices of the feed-forward network (FFN), $\mathbf{W}^{\mathtt{down}}$. \textit{Feature outliers} concentrate in the block output vectors and the input vectors of the down-projection matrices, $\mathbf{x}_{\mathtt{down}}$~\cite{sun2024massive}. Both types of outliers are strongly correlated and persistent in that they are almost always located in the same feature dimension across different layers~\cite{an2025systematic}. Outliers typically emerge in the second block and gradually vanish in the final blocks of a model. Outliers are disproportionately important to model performance~\cite{yin2024outlier}. According to a study on Llama-7B, suppressing only six top outliers has a greater impact than pruning hundreds of thousands of normal weights~\cite{yu2024super}.

\paragraph{Threat Model.}
Our adversary model is similar to those in previous work on hardware-based model protection.
The adversary could be a collusion of malicious end-users, cloud service providers, and network service providers. Its objective is to use the model without the permission of the IP owner or launch a white-box attack against the model. The adversary can \textit{i)} directly access the model architecture and all the model parameters; \textit{ii)} query the \textit{oracle model} through a cloud API or an activated AI hardware; \textit{iii)} obtain a small portion of the training dataset. However, the adversary cannot read or probe the secret key stored in the AI hardware, nor can they run an unauthorized model on the AI hardware. In practice, the secret key can be stored in or derived from a tamper-proof memory~\cite{tuyls2006read}, a trusted platform module (TPM)~\cite{nicholas2021secure}, a physical unclonable function (PUF)~\cite{chakraborty2009harpoon}, an FPGA bitstream~\cite{kamali2018lut}, or a camouflaged circuit layout~\cite{li2017provably}. The adversary aims to either infer the correct key or restore the functionality of the generative model without knowing the correct key.

\section{Locking Methodology}
\label{sec:locking}

\subsection{Overview}
\label{sec:overview}

\begin{figure}[t]
\includegraphics[width=0.47\textwidth]{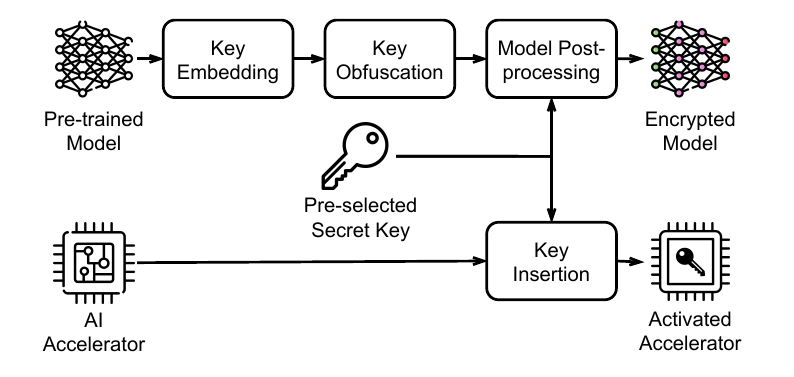}
\caption{Workflow of the LLA model locking framework.}
\label{fig:flow_1}
\end{figure}

\begin{figure}[t]
\hspace*{-12pt}
\includegraphics[width=0.51\textwidth]{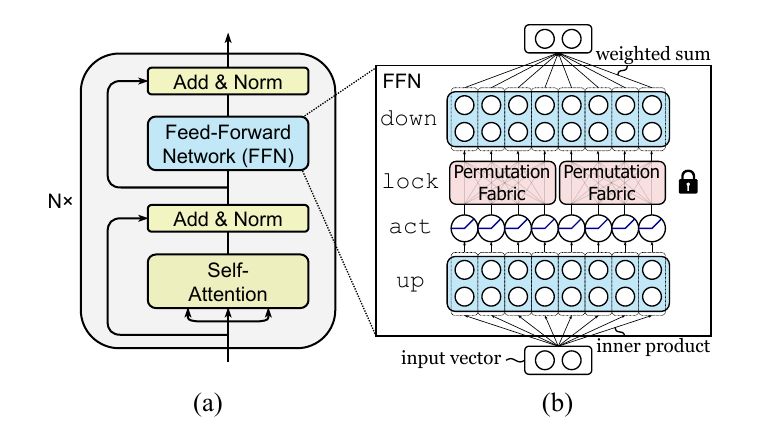}
\caption{(a) Architecture of a transformer block. (b) Simplified illustration of the proposed locking mechanism.
LLA embeds key bits into permutation modules inserted before the down-projection layer, which shuffle intermediate values to alter the model’s functionality.}
\label{fig:flow_2}
\end{figure}

LLA is a post-training model locking approach. Unlike existing approaches such as HPNN~\cite{chakraborty2020hardware}, it does not require additional training resources and can be applied to existing models. LLA aims to achieve the following objectives simultaneously:\\
\noindent $\bullet$ \textit{Effectiveness:} A wrong key will substantially degrade the performance of the GenAI model.\\
\noindent $\bullet$ \textit{Efficiency:} The locking scheme introduces a small number of key bits and incurs negligible performance, power and area overhead.\\
\noindent $\bullet$ 
\textit{Robustness:} An adversary cannot decrypt or bypass the secret key to restore the performance of the model.\\
\noindent $\bullet$ \textit{Hardware friendliness:} The locking scheme can be easily adapted to general AI hardware and requires minimal modifications to both the hardware architecture and the compiler.

The general workflow of the proposed LLA framework is shown in Fig.~\ref{fig:flow_1}. LLA exploits the synergy of software and hardware to achieve these objectives.
At the software level, LLA embeds key bits in the intermediate layer of the FFN module (\cref{sec:embedding}). To maximize corruption to model performance, it selects protected neurons by tracking feature outliers of the model. Afterwards, it employs obfuscation techniques dedicated to GenAI models (\cref{sec:obfuscation}) to protect them from oracle-guided attacks and fine-tuning attacks. Finally, LLA adjusts the model parameters according to the key values. The resulting encrypted model can be sent to the user through a public channel and is functional only with a correct key.
At the hardware level (\cref{sec:hardware}), LLA attaches locking units to the output of the systolic array of an AI accelerator. With a lightweight control unit, LLA dispatches the protected neurons on the fly and triggers the locking units when they reach the output of the systolic array. As such, it does not depend on specific AI compilers and can be easily integrated with most AI accelerators. An accelerator with the inserted secret key serves as a license to access the service provided by the IP owner.

\subsection{Key Embedding}
\label{sec:embedding}

\paragraph{Locking Mechanism.}
LLA designates a transformer block as the \textit{protected block} and selects a subset of neurons within its FFN module as \textit{protected neurons}. The selected neurons are divided into groups with size $m$, and the post-activation values within each group are shuffled using a key-controlled \textit{permutation fabric}. Figure~\ref{fig:flow_2} illustrates the locking mechanism of LLA, which offers three key advantages.
\textit{i)} Localized permutation is agnostic to the input data type and can be efficiently implemented in hardware. Programmable Array Logic (PAL)~\cite{takhar2022holl} and embedded FPGA (eFPGA)~\cite{tang2019openfpga} can be configured to implement interconnections, while butterfly networks~\cite{benevs1964permutation} can realize arbitrary permutations based on control bits.
\textit{ii)} Shuffling an FFN's intermediate values can lead to a more substantial degradation in model performance. Notably, FFN modules act as crucial memory elements in GenAI models~\cite{geva2021transformer}, and the same FFN module is applied repeatedly to all tokens in an input sequence.
\textit{iii)} Placing all key bits within a single block can enhance resilience against a broad spectrum of attacks. Specifically, the discrete nature and high redundancy of the permutation fabric enhance its resilience to gradient-based key decryption attacks. \cref{sec:security} provides a detailed security analysis of LLA.

\paragraph{Identify Protected Neurons.}
GenAI models contain billions of parameters, making them inherently robust to random perturbations. As a result, altering a small subset of randomly selected parameters has a minimal effect on the model outputs.
Recent studies reveal that a small number of outliers are disproportionally important to model performance. In the following, we present a lightweight approach for selecting protected neurons within the FFN module of a designated transformer block.
Intuitively, a neuron can have a significant impact if it can trigger weight outliers within $\mathbf{W}^{\mathtt{down}}$, which then propagate to the hidden state space and produce feature outliers. We trace outliers in reverse to find such critical neurons.
In the first step, we randomly generate a batch of input samples and perform forward passes through the model to identify feature outliers within the output vector of the FFN module:
\begin{equation}
\mathnormal{O}_{f} = \{i \ \vert \  \bar{y}_{i}  > \tau \cdot \mu_{y}\},
\end{equation}
where $\bar{y}_i$ is the average magnitude of the $i$-th feature produced by the designated FFN module, $\mu_{y} \triangleq \frac{1}{D_{\mathrm{m}}}\sum_{i=1}^{D_{\mathrm{m}}} \bar{y}_i$ is the mean of these average magnitudes, and $\tau$ is the threshold that controls the number of selected features.
In the second step, we rank the impact of individual neurons based on the following scoring function:
\begin{equation}
\mathnormal{s}_{j} = \textstyle \sum_{i \in \mathnormal{O}_{f}} \ \lvert \mathbf{W}^{\mathtt{down}}_{j,i} \rvert \cdot \bar{u}_{j},
\end{equation}
where $\bar{u}_{j}$ is the average magnitude of the post-activation value of the $j$-th neuron.
$\mathnormal{s}_{j}$ estimates the contribution of the $j$-th neuron to the features selected in the first step.
Neurons with the highest scores are selected as candidate protected neurons.

As discussed in \cref{sec:background}, feature outliers tend to persist at the same location across consecutive transformer blocks. To prevent information leakage from earlier blocks, we choose the first block exhibiting emergent feature outliers as the protected block.

\subsection{Key Obfuscation}
\label{sec:obfuscation}

Outlier features are disproportionally important for model performance. Therefore, an adversary can launch an approximate oracle-guided attack to recover most of the model's functionality. 
For example, it can identify a small subset of \textit{critical neurons} by iteratively flipping or muting each protected neuron and measuring the impact on the model's output. 
Subsequently, it enumerates all permutations of the subset and selects the one that minimizes the discrepancy with the oracle model~\cite{li2024evaluating}.
In this way, the adversary can quickly find an approximately correct permutation, as the remaining neurons have only a negligible effect on the overall model performance.

We propose an obfuscation method to address this issue. Our technique aims to \textit{i)} smooth the features in the intermediate layer of the protected block, so that each key bit has a similar impact on model performance; \textit{ii)} enhance the correlations among the key bits, so that the quality of model outputs
depends on as many key bits as possible.
LLA utilizes a sequence of orthogonal transformations~\cite{ashkboos2024quarot,lin2024duquant,ashkboosslicegpt} to realize these goals. An orthogonal matrix $\mathbf{M}$ is a square matrix such that $\mathbf{MM^\top} = \mathbf{I}$. Hence, an orthogonal matrix and its transpose can be inserted before and after a sequence of linear layers without changing the functionality of the model. 
We present the details of our obfuscation method in the remainder of this subsection.

\paragraph{Permutation of Outliers.}
In the first step, we build an orthogonal \textit{permutation matrix} $\mathbf{P}$ to reorder neurons within the designated FFN module. Concretely, 
\begin{equation}
\mathbf{P} = \mathbf{P}_1 \cdots \mathbf{P}_n,
\end{equation}
where $\mathbf{P}_j$ swaps the $j$-th protected neuron (\cref{sec:embedding}) with the $j$-th neuron in the original FFN module. After this step, the identified outlier features are repositioned to the front of the feature dimension.

\begin{figure}[t]
\centering

\begin{subfigure}[b]{0.23\textwidth}
    \centering
    \includegraphics[width=\textwidth, trim=0 35 0 40, clip]{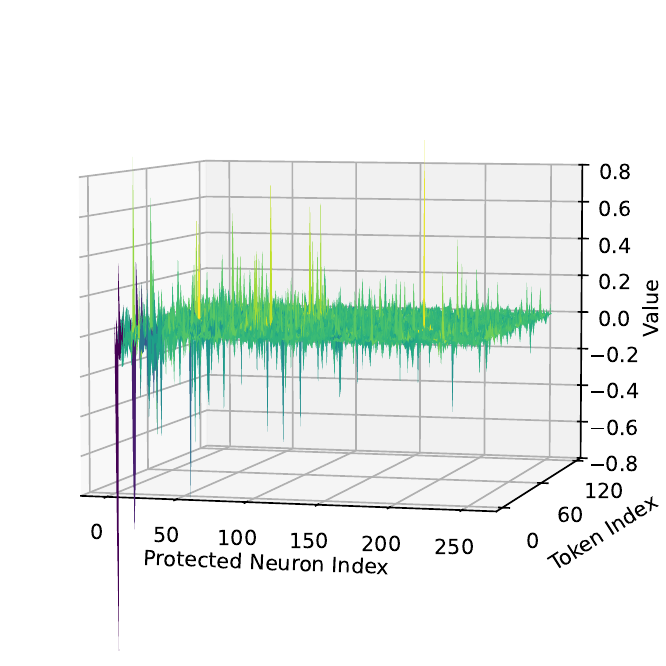}
    \caption{}
    \label{fig:sub1}
\end{subfigure}
\begin{subfigure}[b]{0.23\textwidth}
    \centering
    \includegraphics[width=\textwidth, trim=0 35 0 40, clip]{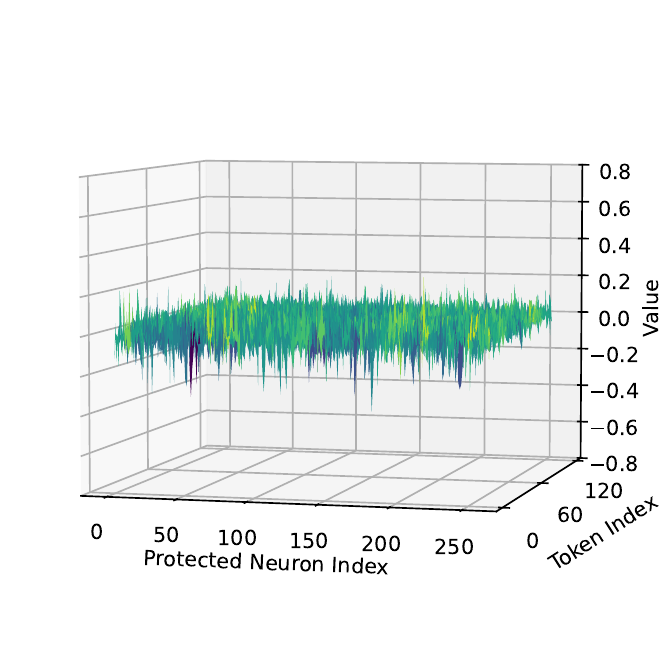}
    \caption{}
    \label{fig:sub2}
\end{subfigure}

\caption{
(a) Feature outliers are prominent before the application of $\mathbf{R}$; (b) Feature outliers are eliminated after the application of $\mathbf{R}$. 
}
\label{fig:rotation}
\end{figure}

\paragraph{Rotation of Features.} 
In the second step, we construct an orthogonal \textit{rotation matrix} $\mathbf{R}$ as follows:
\begin{equation}
\mathbf{R} =
\begin{bmatrix}
    \mathbf{H} & \mathbf{0} \\
    \mathbf{0} & \mathbf{I}
\end{bmatrix}
,
\label{eqn:rotation}
\end{equation}
where $\mathbf{H}$ is a randomized Hadamard matrix~\cite{tseng2024quip,ashkboos2024quarot} and $\mathbf{I}$ is an identity matrix. $\mathbf{H}$ is constructed by scaling an $n$-dimensional Hadamard matrix with $\frac{1}{\sqrt{n}}$, then multiplying it with a \textit{random diagonal matrix} with entries independently sampled from $\{-1,1\}$. A Hadamard matrix is a square matrix whose entries are either $-1$ or $1$ and whose rows are mutually orthogonal.
An $n$-dimensional Hadamard matrix is guaranteed to exist if $n$ is a power of 2, and it is known to exist for almost all $n$ that is a multiple of 4 and less than 1000~\cite{wallis1976existence}.
$\mathbf{R}$ is inserted after the activation layer for a standard FFN module, or after the element-wise multiplication operator for a gated FFN module. 
Fig.~\ref{fig:rotation} visualizes the smoothing effect of the rotation matrix.

The rotation transformation has the following benefits: \textit{i)} it evenly distributes the effects of outlier features across the first $n$ dimensions by smoothing their magnitudes; \textit{ii)} it enhances the interdependence among key bits, thereby obscuring the statistical relationships between key bits and model outputs~\cite{shannon1949communication}; \textit{iii)} it discretizes the effects of outlier features in the output space of $\mathbf{R}$, thus mitigating gradient-based oracle-guided attacks.

\paragraph{Insertion of Keys.} In the third step, we create an orthogonal \textit{key matrix} $\mathbf{K}$ given a permutation $\pi$ that rearrange $n$ elements:
\begin{equation}
\mathbf{K} =
\begin{bmatrix}
    \mathbf{G} & \mathbf{0} \\
    \mathbf{0} & \mathbf{I}
\end{bmatrix}
,
\end{equation}
where $\mathbf{G}$ is also a permutation matrix. Each element $\mathbf{G}_{i,j}$ is equal to $1$ if $\boldsymbol{\pi}(i) = j$, and $0$ otherwise. To reduce hardware complexity, we further require that $\boldsymbol{\pi}$ can be partitioned into disjoint groups, each of which is a permutation of $m$ elements. As a result, the computation of $\mathbf{G}$ can be realized using key-controlled permutation fabrics of size $m$.
We elaborate on the hardware support in \cref{sec:hardware}.

\paragraph{Obfuscation with Orthogonal Transformations.} 
A standard FFN module can be expressed as $\mathbf{Y} = \sigma(\mathbf{X}\mathbf{W}^{\mathtt{up}}) \mathbf{W}^{\mathtt{down}}$, where $\sigma$ denotes the element-wise activation function, $\mathbf{X}$ represents the input matrix of the FFN module, and $\mathbf{Y}$ represents the corresponding output matrix.
The overall obfuscation method can be summarized as the following:
\begin{equation}
\mathbf{Y} = \sigma ( \mathbf{X} \mathbf{W}^{\mathtt{up}} \mathbf{P} )
\mathbf{R} \mathbf{K}
( \mathbf{K}^\top \mathbf{R}^\top  \mathbf{P}^\top \mathbf{W}^{\mathtt{down}} )
.
\end{equation}

After commuting $\mathbf{P}$ with $\sigma(\cdot)$, each orthogonal matrix is canceled by its transpose, so the final output after transformations remains unchanged. 
To reduce computational cost, we merge all orthogonal matrices within the same parentheses into $\mathbf{W}^{\mathtt{up}}$ or $\mathbf{W}^{\mathtt{down}}$, resulting in the following transformations:
\begin{equation}
\mathbf{Y} = \sigma ( \mathbf{X} \tilde{\mathbf{W}}^{\mathtt{up}} )
\mathbf{R} \mathbf{K}
( \tilde{\mathbf{W}}^{\mathtt{down}} )
.
\label{eqn:simplified}
\end{equation}

As shown in Formula~\ref{eqn:rotation}, the main component of $\mathbf{R}$ is a randomized Hadamard matrix $\mathbf{H}$, whose dimensionality corresponds to the key size $n$. 
In contrast, the dimensionality of $\mathbf{R}$ matches the intermediate dimension $D_{\mathrm{ff}}$ of the FFN module, which is typically on the order of 10,000 or higher in modern GenAI models~\cite{dubey2024llama}. Therefore, when the key size is on the order of hundreds, the computational overhead caused by $\mathbf{R}$ is negligible. 
Notice that the random diagonal matrix within $\mathbf{R}^\top$ introduces additional confusion to $\tilde{\mathbf{W}}^{\mathtt{down}}$, helping to offset the potential information leakage caused by $\mathbf{K}^\top$.

The above method can also be applied to a gated FFN module by distributing $\mathbf{P}$ to both $\mathbf{W}^\mathtt{up}$ and $\mathbf{W}^\mathtt{gate}$.

\subsection{Hardware Support}
\label{sec:hardware}

\begin{figure}[t]
\centering
\includegraphics[width=0.33\textwidth, trim=10 30 10 30, clip]{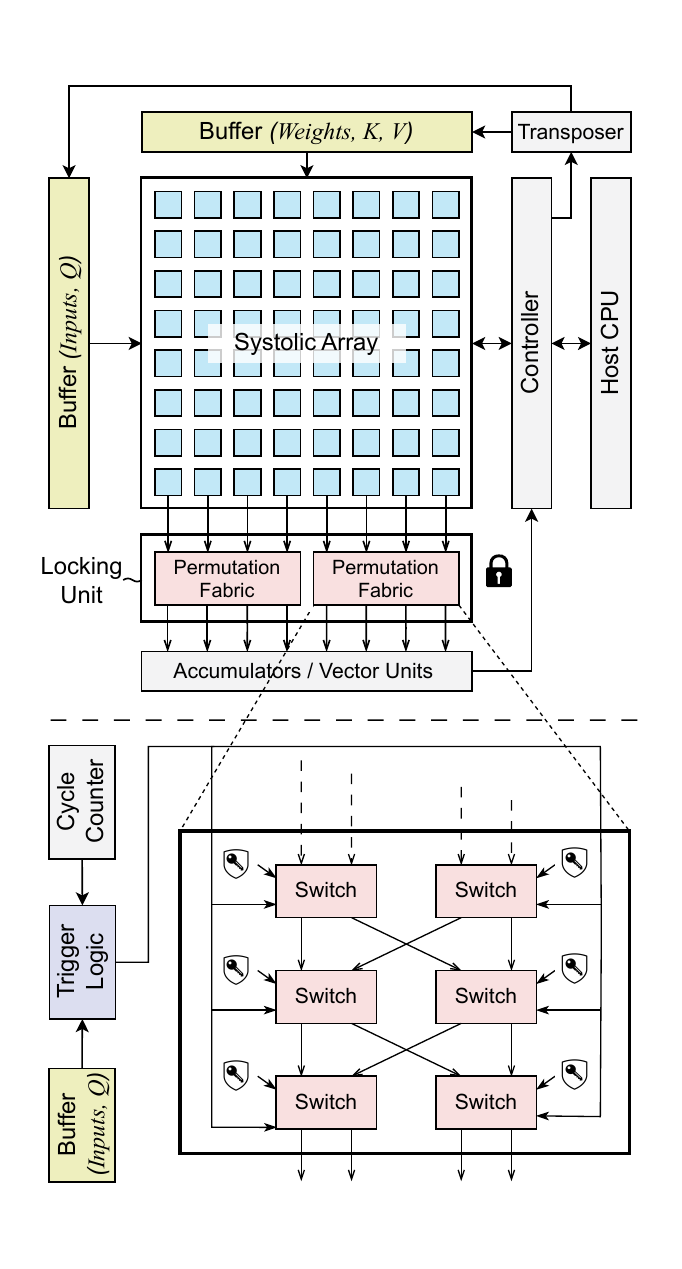}
\caption{Schematic of a systolic AI accelerator that supports model locking.}
\label{fig:spatial}
\end{figure}

\paragraph{Systolic Array Architecture.} 
The \textit{systolic array} is the core computational module in modern AI accelerators~\cite{chen2020survey,ju202265nm}. It comprises a mesh of interconnected process elements (PEs), each performing a scalar multiplication and accumulation (MAC) operation in a clock cycle. \textit{Weight-stationary} and \textit{output-stationary} are two representative dataflow schemes of systolic arrays. In the weight-stationary scheme, a tile of the weight matrix is retained with the PEs, whereas the input matrix and the partial sums are streamed through the PEs during computation. In the output-stationary scheme, the partial sums are retained within the PEs, while the input and weight matrices are streamed through the PEs.

\paragraph{Hardware Design.}
LLA is designed to be agnostic to specific AI compilers or accelerators.
Fig.~\ref{fig:spatial}\textit{(left)} illustrates the architecture of a systolic array accelerator that supports model locking.
The \textit{locking module} consists of several permutation fabrics, each shuffling a fixed number of consecutive output lanes of the systolic array.
This additional module is responsible for performing the multiplication with the key matrix $\mathbf{K}$.
Fig.~\ref{fig:spatial}\textit{(right)} shows an implementation of a 4$\times$4 permutation fabric with a key-controlled Benes network~\cite{benevs1964permutation}. The network comprises multiple stages of 2$\times$2 switches, each controlled by a key bit that determines whether to pass through or swap the two input signals. 
The \textit{trigger logic} is inactive when the systolic array computes any matrix other than the key matrix $\mathbf{K}$. In that case, it outputs the default key pattern that preserves the original order of all lanes. When the systolic array computes the $\mathbf{K}$ matrix, the trigger unit introduces specific delays to each output lane based on the dataflow schemes, ensuring that the signals reach the permutation fabric within the same clock cycle. Meanwhile, it configures the routing of the Benes network with input key bits to realize the desired permutation.
Weight-stationary systolic arrays may compute an output matrix in multiple rounds by accumulating several intermediate matrices. LLA applies the same key to each intermediate matrix to achieve the intended result.

\section{Experiment and Analysis}
\label{sec:experiment}

\paragraph{Setup.} 
We apply the proposed LLA model locking approach to four pre-trained LLMs from the Minitron family~\cite{muralidharan2024compact}. These models represent different types of FFNs with varying sizes. Their statistics are summarized in Table~\ref{table:models-to-eval}. 
As model size increases, LLMs tend to exhibit more feature outliers~\cite{dettmers2022gpt3}, which in turn raises the cost of successful attacks~\cite{li2024evaluating}. Consequently, the positive results observed on these relatively small models indicate that LLA can be more effective when applied to larger models.
All experiments are conducted on a Linux workstation with a 2.4 GHz CPU and an NVIDIA RTX A6000 GPU.

We use $\mathtt{MMLU}$ and $\mathtt{perplexity}$ to evaluate the capabilities of LLMs. A higher MMLU score or a lower perplexity indicates better performance.
$\mathtt{Fidelity}$ is defined as the proportion of protected neurons whose original indices are restored by an attacking algorithm.
We choose the Jensen-Shannon divergence ($\mathtt{JSD}$) to measure the similarity between the output distributions of the encrypted model and the oracle model.

The primary threat to model locking is the oracle-guided ($\mathtt{OG}$) attack. In this setting, an adversary can query the oracle model with any input sequence and observe the corresponding output logits.
If the oracle model is unavailable, the adversary can instead launch a pretraining-style oracle-less ($\mathtt{OL}$) attack using a curated dataset. 
We consider two prevalent optimization techniques for key decryption.
The \textit{genetic-based attack}~\cite{alam2022nn} iteratively evolves the candidate key pattern in the binary key space.
Our implementation employs the tournament selection mechanism, uses JSD as the fitness function, applies within-group pairwise swaps as the mutation operator, and performs crossover across permutation groups.
For the \textit{gradient-based attack}~\cite{li2024evaluating}, existing methods cannot be applied directly to the permutation fabric. According to our experiments, the most effective way is to directly restore the permutation matrix $\mathbf{G}$. Specifically, we use a \textit{softmax} function to approximate each column of the matrix. Upon completion, the element with the highest magnitude is set to $1$, and the remaining elements are set to $0$. We use Adam optimization with a learning rate of 0.03. We choose JSD as the loss function for the OG attack and cross-entropy for the OL attack.
We observe that the gradient-based attack is consistently more effective than the genetic-based attack. Therefore, our evaluation focuses primarily on the former attack.
We set a time limit of 7,200 seconds for every execution.

We assess LLA with various numbers of protected neurons, ranging from 256 to 8192. By default, the permutation group size is set to 16. 
We configure the outlier threshold $\tau$ to 5, which can be increased for larger generative models. Based on the emergence of outliers, we select the first block of \texttt{Mistral NeMo Minitron 8B} and the second block of the other three models as protected blocks.

\paragraph{Results.}

\begin{figure}[t]
\centering
\includegraphics[width=0.46\textwidth]{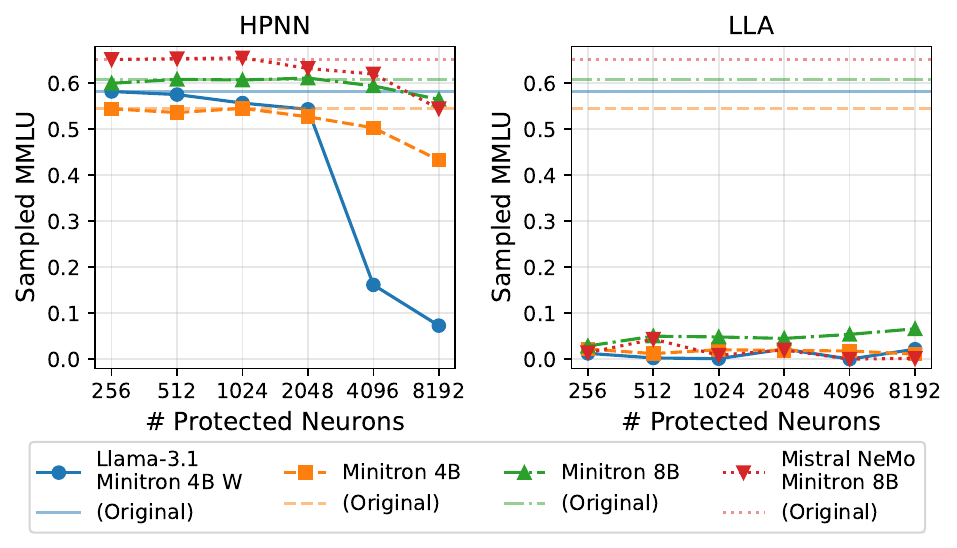}
\caption{Pre-attack locking effectiveness: LLA \textit{vs.} HPNN.
}
\label{fig:experiments-lla-pre-attack-mmu}
\end{figure}

\begin{figure}[t]
\centering
\includegraphics[width=0.46\textwidth]{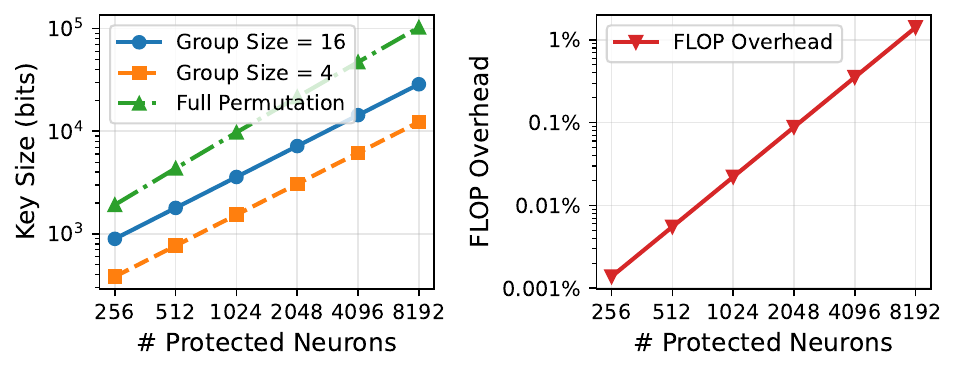}
\caption{
Locking efficiency: key size and FLOP overhead.
}
\label{fig:experiments-lla-overhead}
\end{figure}

\begin{figure}[t]
\centering
\includegraphics[width=0.46\textwidth]{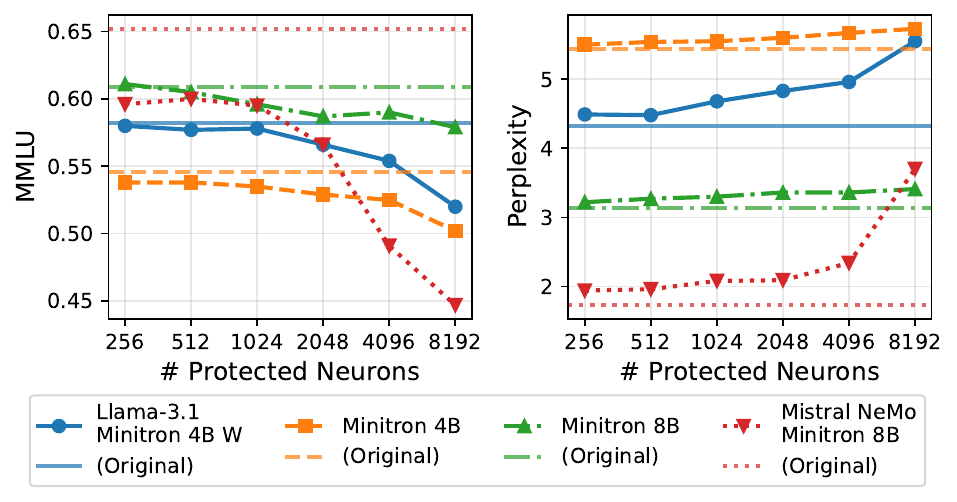}
\caption{Locking robustness: performance of LLA-protected models after the OG gradient-based attack. }
\label{fig:experiments-16-way}
\end{figure}

\begin{figure}[t]
\centering
\includegraphics[width=0.46\textwidth]{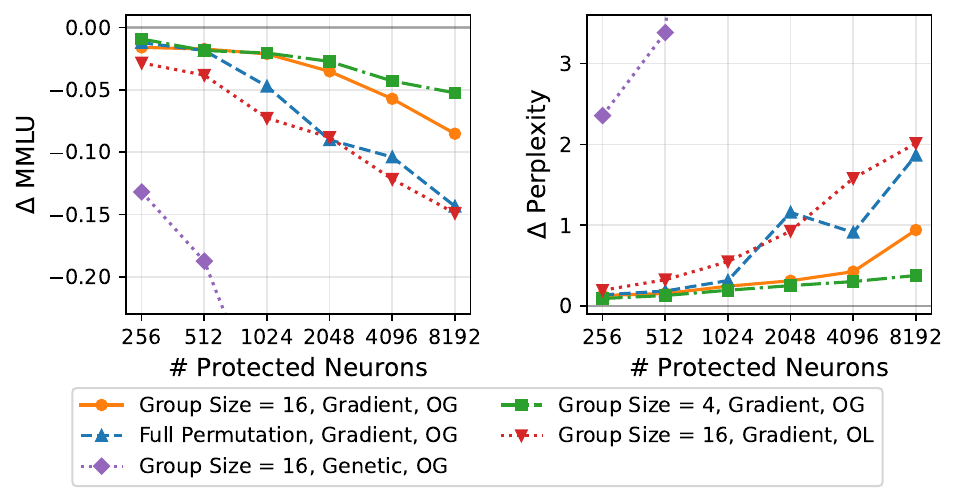}
\caption{
Locking robustness: performance of LLA-protected models under various attack and defense settings
(averaged across 4 models).
}
\label{fig:experiments-lla-variants}
\end{figure}

We systematically evaluate LLA in terms of its effectiveness, efficiency, and robustness. First, we assess the locking effectiveness of LLA by comparing it with HPNN~\cite{chakraborty2020hardware}, the state-of-the-art model locking technique. For each locking configuration, we report the average model performance across 10 randomly generated key patterns. As shown in Figure~\ref{fig:experiments-lla-pre-attack-mmu}, HPNN causes only a slight performance degradation in most configurations. In contrast, LLA consistently renders the models unusable in all configurations.

Second, we measure the locking efficiency of LLA in terms of key length and computational overhead (Figure~\ref{fig:experiments-lla-overhead}). For a fixed permutation group size, the key length grows linearly with the number of protected neurons. LLA requires 3.5 key bits per protected neuron for a group size of 16 and 1.5 bits per neuron for a group size of 4.
On the other hand, the computational overhead in FLOP count is negligible, especially for smaller key sizes.
The effectiveness and efficiency of LLA can be attributed to its ability to manipulate outliers and absorb orthogonal matrices.

\begin{figure}[t]
\centering
\includegraphics[width=0.46\textwidth]{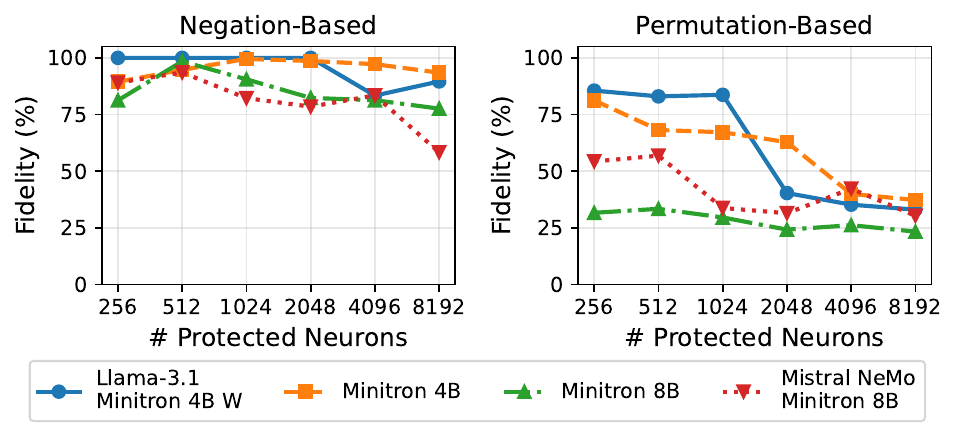}
\caption{
Ablation study: comparing the locking robustness of negation-based and permutation-based locking schemes. 
}
\label{fig:experiments-lla-fidelity}
\end{figure}

Third, we evaluate the locking robustness of LLA by testing it against the aforementioned genetic-based and gradient-based attacks. For each configuration, we report the \textit{best} attack result observed across three attempts. As depicted in Figure~\ref{fig:experiments-16-way}, post-attack performance varies by model type, but robustness consistently improves as the key length increases. 
The fidelity (Figure~\ref{fig:experiments-lla-fidelity}\textit{(right)}) never exceeds 86\% for a small key length (256 protected neurons) and 38\% for a large key length (8192 protected neurons), suggesting that an adversary cannot recover the original functionality of the model.
We further analyze the robustness of LLA under various attack and defense settings (Figure~\ref{fig:experiments-lla-variants}). We observe that \textit{i)} robustness can be enhanced with a greater permutation group size, though it comes at the expense of increased key length; \textit{ii)} the gradient-based attack outperforms the genetic-based attack; \textit{iii)} OG attacks are more effective than OL attacks when other configurations are the same.

\begin{table}[t]
\small
\centering
\begin{tabular}{cccc}
\hline 
\multirow{2}{*}{\textbf{Model}} & \textbf{FFN Type} & \multicolumn{2}{c}{\textbf{Original Perf.}} \\
 & $D_\mathrm{m} \times D_\mathrm{ff}$ & $\ \mathtt{MMLU}\ $ & $\!\!\mathtt{Perplexity}\!\!$ \\
\hline 
\multirow{2}{*}{\texttt{Minitron 4B}} & standard & \multirow{2}{*}{0.546} & \multirow{2}{*}{5.44} \\
 & 3072 $\times$ 9216 &  &  \\
\hline 
\multirow{2}{*}{\texttt{Minitron 8B}} & standard & \multirow{2}{*}{0.609} & \multirow{2}{*}{3.14} \\
 & 4096 $\times$ 16384 &  &  \\
\hline 
\texttt{Llama-3.1} & gated & \multirow{2}{*}{0.582} & \multirow{2}{*}{4.32} \\
\texttt{Minitron 4B W} & 3072 $\times$ 9216 &  &  \\
\hline 
\texttt{Mistral Nemo} & gated & \multirow{2}{*}{0.652} & \multirow{2}{*}{1.73} \\
\texttt{Minitron 8B} & 4096 $\times$ 11520 &  &  \\
\hline 
\end{tabular}
\caption{LLMs used for evaluation.}
\label{table:models-to-eval}
\end{table}

\paragraph{Ablation Study.} 

\begin{figure}[t]
\centering
\includegraphics[width=0.46\textwidth]{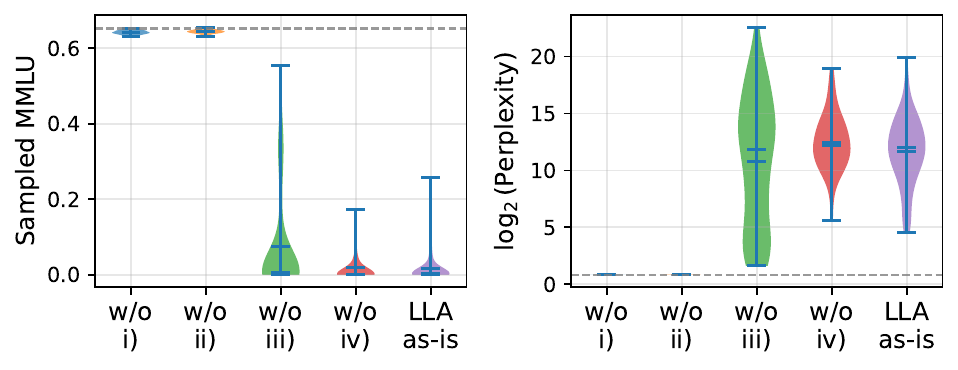}
\caption{
Ablation study: pre-attack performance of \texttt{Mistral NeMo Minitron 8B} under various LLA configurations. The dashed lines indicate the performance of the original model.
}
\label{fig:ablation}
\end{figure}

We conduct an ablation study to identify the key factors that contribute to the effectiveness and robustness of LLA.
Specifically, we examine the following factors: \textit{i)} selection of the protected block, \textit{ii)} selection of the protected neurons, \textit{iii)} application of the rotation matrix, and \textit{iv)} use of permutation-based locking instead of negation-based locking.
We select HPNN as the baseline method.

We compare the locking effectiveness across various LLA configurations in a non-adversarial setting. For each configuration, we randomly generate $100$ key patterns and plot the corresponding model performance in Fig.~\ref{fig:ablation}. Removing either \textit{i)} or \textit{ii)} leads to minimal performance degradation, highlighting the critical role of outlier features. Removing \textit{iii)} still allows certain key patterns to maintain relatively high performance.
Further analysis reveals that these patterns retain the original positions of most critical neurons, exposing the vulnerability of model locking under insufficient obfuscation.
The last two configurations consistently degrade model performance across all key patterns, demonstrating their effectiveness in a non-adversarial setting.

Finally, we evaluate the locking robustness of the last two configurations in Fig.~\ref{fig:ablation}. The negation-based locking scheme was first proposed by HPNN, and the permutation-based locking scheme was introduced by LLA. For the negation-based locking scheme, we replace every key bit with a \textit{tanh} function during the attack. Upon convergence, the key bits with negative values are set to $\mathtt{true}$, while the remaining bits are set to $\mathtt{false}$~\cite{li2024evaluating}.
Figure~\ref{fig:experiments-lla-fidelity} compares the post-attack fidelity of the two schemes. The negation-based scheme is more vulnerable to gradient-based attacks. Notably, the adversary can even recover 100\% of key bits for the \texttt{Llama-3.1 Minitron 4B W} model under various settings. Therefore, the permutation-based scheme is necessary to ensure the robustness of model locking.

\paragraph{Security Analysis.}
\label{sec:security}

As discussed previously, LLA-protected models are resistant to \textit{oracle-guided key optimization attacks}. 
Alternatively, an adversary may fix a key value and launch \textit{oracle-guided fine-tuning attacks} to improve model performance. 
However, training with an incorrectly fixed key can be more costly than training from scratch.
LLA is also resistant to \textit{probing attacks} and various \textit{side-channel attacks} because: \textit{i)} there are only delay registers between the systolic array and the locking module, and \textit{ii)} the execution of a protected model follows a fixed pattern that is independent of the key value.
The adversary may exploit the geometric and algebraic properties of neural networks to launch \textit{oracle-guided geometric attacks}~\cite{li2024evaluating}. 
LLA can thwart these attacks because: \textit{i)} all key bits are placed within the same intermediate layer, which has a larger dimension $D_{\mathrm{ff}}$ than the input dimension $D_{\mathrm{m}}$; geometric attacks are ineffective on expansive layers. \textit{ii)} The key bits are tightly correlated due to orthogonal transformations, whereas geometric attacks rely on a divide-and-conquer strategy to reduce computational complexity.

\section{Conclusion and Future Work}
\label{sec:conclusion}

This paper presents LLA, the first model locking method tailored for large generative models. It brings together a set of techniques, including outlier selection, key obfuscation, and systolic array hardware support, to safeguard the supply chain of generative models. Experiments demonstrate that LLA can mitigate a wide range of attacks, particularly oracle-guided key optimization attacks, at a minimal computational and hardware cost.
Limitations of this work include: \textit{i)} it is not as effective on tiny generative models where prominent outliers are absent; \textit{ii)} an adversary with sufficient computational resources may leverage model distillation to replace the entire protected block. We plan to address these challenges in future work.

\section*{Acknowledgments}
This work is partially supported by the National Science Foundation under grants 2113704 and 2148177.

\bibliography{ref}

@inproceedings{zhuang2023pilot,
  title={A pilot study of query-free adversarial attack against stable diffusion},
  author={Zhuang, Haomin and Zhang, Yihua and Liu, Sijia},
  booktitle={CVPR 2023},
  year={2023}
}

@article{zou2023universal,
  title={Universal and transferable adversarial attacks on aligned language models},
  author={Zou, Andy and Wang, Zifan and Carlini, Nicholas and Nasr, Milad and Kolter, J Zico and Fredrikson, Matt},
  journal={arXiv:2307.15043},
  year={2023}
}

@inproceedings{zhao2023prompt,
  title={Prompt as Triggers for Backdoor Attack: Examining the Vulnerability in Language Models},
  author={Zhao, Shuai and Wen, Jinming and Luu, Anh and Zhao, Junbo and Fu, Jie},
  booktitle={EMNLP 2023},
  year={2023}
}

@inproceedings{chou2023backdoor,
  title={How to backdoor diffusion models?},
  author={Chou, Sheng-Yen and Chen, Pin-Yu and Ho, Tsung-Yi},
  booktitle={CVPR 2023},
  year={2023}
}

@article{kamali2022advances,
  title={Advances in logic locking: Past, present, and prospects},
  author={Kamali, Hadi Mardani and Azar, Kimia Zamiri and Farahmandi, Farimah and Tehranipoor, Mark},
  journal={Future Microelectronics Security Research Series},
  year={2022}
}

@article{nicholas2021secure,
  title={A secure boot framework with multi-security features and logic-locking applications for reconfigurable logic},
  author={Nicholas, Geraldine Shirley and Siddiqui, Ali Shuja and Joseph, Sam Reji and Williams, Gregory and Saqib, Fareena},
  journal={Journal of Hardware and Systems Security},
  volume={5},
  number={3},
  pages={260--268},
  year={2021},
  publisher={Springer}
}

@article{chakraborty2009harpoon,
  title={HARPOON: An obfuscation-based SoC design methodology for hardware protection},
  author={Chakraborty, Rajat Subhra and Bhunia, Swarup},
  journal={IEEE Transactions on Computer-Aided Design of Integrated Circuits and Systems},
  volume={28},
  number={10},
  pages={1493--1502},
  year={2009}
}

@inproceedings{kamali2018lut,
  title={Lut-lock: A novel lut-based logic obfuscation for fpga-bitstream and asic-hardware protection},
  author={Kamali, Hadi Mardani and Azar, Kimia Zamiri and Gaj, Kris and Homayoun, Houman and Sasan, Avesta},
  booktitle={ISVLSI 2018},
  year={2018}
}

@inproceedings{chakraborty2020hardware,
  title={Hardware-assisted intellectual property protection of deep learning models},
  author={Chakraborty, Abhishek and Mondai, Ankit and Srivastava, Ankur},
  booktitle={DAC 2020},
  year={2020}
}

@article{alam2022nn,
  title={NN-Lock: A lightweight authorization to prevent IP threats of deep learning models},
  author={Alam, Manaar and Saha, Sayandeep and Mukhopadhyay, Debdeep and Kundu, Sandip},
  journal={ACM Journal on Emerging Technologies in Computing Systems (JETC)},
  volume={18},
  number={3},
  pages={1--19},
  year={2022}
}

@article{goldstein2021preventing,
  title={Preventing DNN model IP theft via hardware obfuscation},
  author={Goldstein, Brunno F and Patil, Vinay C and Ferreira, Victor C and Nery, Alexandre S and Fran{\c{c}}a, Felipe MG and Kundu, Sandip},
  journal={IEEE Journal on Emerging and Selected Topics in Circuits and Systems},
  volume={11},
  number={2},
  pages={267--277},
  year={2021}
}

@inproceedings{li2024evaluating,
  title={Evaluating the Security of Logic Locking on Deep Neural Networks},
  author={Li, You and Zhao, Guannan and He, Yunqi and Zhou, Hai},
  booktitle={DAC 2024},
  year={2024}
}

@inproceedings{kirchenbauer2023watermark,
  title={A watermark for large language models},
  author={Kirchenbauer, John and Geiping, Jonas and Wen, Yuxin and Katz, Jonathan and Miers, Ian and Goldstein, Tom},
  booktitle={ICML 2023},
  year={2023}
}

@article{xu2024instructional,
  title={Instructional fingerprinting of large language models},
  author={Xu, Jiashu and Wang, Fei and Ma, Mingyu Derek and Koh, Pang Wei and Xiao, Chaowei and Chen, Muhao},
  journal={arXiv:2401.12255},
  year={2024}
}

@inproceedings{mo2020darknetz,
  title={Darknetz: towards model privacy at the edge using trusted execution environments},
  author={Mo, Fan and Shamsabadi, Ali Shahin and Katevas, Kleomenis and Demetriou, Soteris and Leontiadis, Ilias and Cavallaro, Andrea and Haddadi, Hamed},
  booktitle={MobiSys 2020},
  year={2020}
}

@article{lin2020chaotic,
  title={Chaotic weights: A novel approach to protect intellectual property of deep neural networks},
  author={Lin, Ning and Chen, Xiaoming and Lu, Hang and Li, Xiaowei},
  journal={IEEE Transactions on Computer-Aided Design of Integrated Circuits and Systems},
  volume={40},
  number={7},
  pages={1327--1339},
  year={2020}
}

@inproceedings{zuo2021sealing,
  title={Sealing neural network models in encrypted deep learning accelerators},
  author={Zuo, Pengfei and Hua, Yu and Liang, Ling and Xie, Xinfeng and Hu, Xing and Xie, Yuan},
  booktitle={DAC 2021},
  year={2021}
}

@inproceedings{tuyls2006read,
  title={Read-proof hardware from protective coatings},
  author={Tuyls, Pim and Schrijen, Geert-Jan and {\v{S}}kori{\'c}, Boris and Van Geloven, Jan and Verhaegh, Nynke and Wolters, Rob},
  booktitle={8th International Workshop of Cryptographic Hardware and Embedded Systems, CHES 2006},
  year={2006}
}

@article{li2017provably,
  title={Provably secure camouflaging strategy for IC protection},
  author={Li, Meng and Shamsi, Kaveh and Meade, Travis and Zhao, Zheng and Yu, Bei and Jin, Yier and Pan, David Z},
  journal={IEEE transactions on computer-aided design of integrated circuits and systems},
  volume={38},
  number={8},
  pages={1399--1412},
  year={2017}
}

@inproceedings{ju202265nm,
  title={A 65nm systolic neural CPU processor for combined deep learning and general-purpose computing with 95\% PE utilization, high data locality and enhanced end-to-end performance},
  author={Ju, Yuhao and Gu, Jie},
  booktitle={ISSCC 2022},
  year={2022}
}

@inproceedings{geva2021transformer,
  title={Transformer Feed-Forward Layers Are Key-Value Memories},
  author={Geva, Mor and Schuster, Roei and Berant, Jonathan and Levy, Omer},
  booktitle={EMNLP 2021},
  year={2021}
}

@article{men2024shortgpt,
  title={Shortgpt: Layers in large language models are more redundant than you expect},
  author={Men, Xin and Xu, Mingyu and Zhang, Qingyu and Wang, Bingning and Lin, Hongyu and Lu, Yaojie and Han, Xianpei and Chen, Weipeng},
  journal={arXiv:2403.03853},
  year={2024}
}

@inproceedings{sun2024massive,
  title={Massive Activations in Large Language Models},
  author={Sun, Mingjie and Chen, Xinlei and Kolter, J Zico and Liu, Zhuang},
  booktitle={ICLR 2024 Workshop on Mathematical and Empirical Understanding of Foundation Models},
  year={2024}
}

@article{dettmers2022gpt3,
  title={Gpt3. int8 (): 8-bit matrix multiplication for transformers at scale},
  author={Dettmers, Tim and Lewis, Mike and Belkada, Younes and Zettlemoyer, Luke},
  journal={NIPS 2022},
  year={2022}
}

@article{yu2024super,
  title={The Super Weight in Large Language Models},
  author={Yu, Mengxia and Wang, De and Shan, Qi and Wan, Alvin},
  journal={arXiv:2411.07191},
  year={2024}
}

@inproceedings{lilazy,
  title={The Lazy Neuron Phenomenon: On Emergence of Activation Sparsity in Transformers},
  author={Li, Zonglin and You, Chong and Bhojanapalli, Srinadh and Li, Daliang and Rawat, Ankit Singh and Reddi, Sashank J and Ye, Ke and Chern, Felix and Yu, Felix and Guo, Ruiqi and others},
  booktitle={ICLR 2023},
  year={2023}
}

@article{dubey2024llama,
  title={The llama 3 herd of models},
  author={Dubey, Abhimanyu and Jauhri, Abhinav and Pandey, Abhinav and Kadian, Abhishek and Al-Dahle, Ahmad and Letman, Aiesha and Mathur, Akhil and Schelten, Alan and Yang, Amy and Fan, Angela and others},
  journal={arXiv:2407.21783},
  year={2024}
}

@inproceedings{uchida2017embedding,
  title={Embedding watermarks into deep neural networks},
  author={Uchida, Yusuke and Nagai, Yuki and Sakazawa, Shigeyuki and Satoh, Shin'ichi},
  booktitle={Proceedings of the 2017 ACM on international conference on multimedia retrieval},
  year={2017}
}

@inproceedings{adi2018turning,
  title={Turning your weakness into a strength: Watermarking deep neural networks by backdooring},
  author={Adi, Yossi and Baum, Carsten and Cisse, Moustapha and Pinkas, Benny and Keshet, Joseph},
  booktitle={USENIX Security 2018},
  year={2018}
}

@inproceedings{zhou2023nnsplitter,
  title={NNSplitter: an active defense solution for DNN model via automated weight obfuscation},
  author={Zhou, Tong and Luo, Yukui and Ren, Shaolei and Xu, Xiaolin},
  booktitle={ICML 2023},
  year={2023}
}

@inproceedings{mu2024encryip,
  title={EncryIP: A Practical Encryption-Based Framework for Model Intellectual Property Protection},
  author={Mu, Xin and Wang, Yu and Huang, Zhengan and Lai, Junzuo and Zhang, Yehong and Wang, Hui and Yu, Yue},
  booktitle={AAAI 2024},
  year={2024}
}

@article{li2025licensenet,
  title={LicenseNet: Proactively safeguarding intellectual property of AI models through model license},
  author={Li, Peihao and Huang, Jie and Zhang, Shuaishuai},
  journal={Journal of Systems Architecture},
  volume={159},
  pages={103330},
  year={2025},
  publisher={Elsevier}
}

@article{khan2021utilizing,
  title={Utilizing and extending trusted execution environment in heterogeneous SoCs for a pay-per-device IP licensing scheme},
  author={Khan, Nadir and Nitzsche, Sven and L{\'o}pez, Asier Garciandia and Becker, J{\"u}rgen},
  journal={IEEE Transactions on Information Forensics and Security},
  volume={16},
  pages={2548--2563},
  year={2021},
  publisher={IEEE}
}

@article{wang2023building,
  title={Building a lightweight trusted execution environment for arm gpus},
  author={Wang, Chenxu and Deng, Yunjie and Ning, Zhenyu and Leach, Kevin and Li, Jin and Yan, Shoumeng and He, Zhengyu and Cao, Jiannong and Zhang, Fengwei},
  journal={IEEE Transactions on Dependable and Secure Computing},
  volume={21},
  number={4},
  pages={3801--3816},
  year={2023},
  publisher={IEEE}
}

@inproceedings{gilad2016cryptonets,
  title={Cryptonets: Applying neural networks to encrypted data with high throughput and accuracy},
  author={Gilad-Bachrach, Ran and Dowlin, Nathan and Laine, Kim and Lauter, Kristin and Naehrig, Michael and Wernsing, John},
  booktitle={ICML 2016},
  year={2016}
}

@article{sun2018private,
  title={Private machine learning classification based on fully homomorphic encryption},
  author={Sun, Xiaoqiang and Zhang, Peng and Liu, Joseph K and Yu, Jianping and Xie, Weixin},
  journal={IEEE Transactions on Emerging Topics in Computing},
  volume={8},
  number={2},
  pages={352--364},
  year={2018},
  publisher={IEEE}
}

@article{kovaleva2021bert,
  title={Bert busters: Outlier dimensions that disrupt transformers},
  author={Kovaleva, Olga and Kulshreshtha, Saurabh and Rogers, Anna and Rumshisky, Anna},
  journal={arXiv:2105.06990},
  year={2021}
}

@article{an2025systematic,
  title={Systematic Outliers in Large Language Models},
  author={An, Yongqi and Zhao, Xu and Yu, Tao and Tang, Ming and Wang, Jinqiao},
  booktitle={ICLR 2025},
  year={2025}
}

@inproceedings{yin2024outlier,
  title={Outlier Weighed Layerwise Sparsity (OWL): A Missing Secret Sauce for Pruning LLMs to High Sparsity},
  author={Yin, Lu and Wu, You and Zhang, Zhenyu and Hsieh, Cheng-Yu and Wang, Yaqing and Jia, Yiling and Li, Gen and Jaiswal, Ajay and Pechenizkiy, Mykola and Liang, Yi and others},
  booktitle={ICML 2024},
  year={2024}
}

@inproceedings{ashkboosslicegpt,
  title={SliceGPT: Compress Large Language Models by Deleting Rows and Columns},
  author={Ashkboos, Saleh and Croci, Maximilian L and do Nascimento, Marcelo Gennari and Hoefler, Torsten and Hensman, James},
  journal={ICLR 2024},
  year={2024}
}

@article{wallis1976existence,
  title={On the existence of Hadamard matrices},
  author={Wallis, Jennifer Seberry},
  journal={Journal of Combinatorial Theory, Series A},
  volume={21},
  number={2},
  pages={188--195},
  year={1976},
  publisher={Elsevier}
}

@inproceedings{tseng2024quip,
  title={QuIP $\# $: Even Better LLM Quantization with Hadamard Incoherence and Lattice Codebooks},
  author={Tseng, Albert and Chee, Jerry and Sun, Qingyao and Kuleshov, Volodymyr and De Sa, Christopher},
  journal={ICML 2024},
  year={2024}
}

@article{shannon1949communication,
  title={Communication theory of secrecy systems},
  author={Shannon, Claude E},
  journal={The Bell system technical journal},
  volume={28},
  number={4},
  pages={656--715},
  year={1949},
  publisher={Nokia Bell Labs}
}

@article{ashkboos2024quarot,
  title={Quarot: Outlier-free 4-bit inference in rotated llms},
  author={Ashkboos, Saleh and Mohtashami, Amirkeivan and Croci, Maximilian and Li, Bo and Cameron, Pashmina and Jaggi, Martin and Alistarh, Dan and Hoefler, Torsten and Hensman, James},
  journal={NIPS 2024},
  year={2024}
}

@article{lin2024duquant,
  title={Duquant: Distributing outliers via dual transformation makes stronger quantized llms},
  author={Lin, Haokun and Xu, Haobo and Wu, Yichen and Cui, Jingzhi and Zhang, Yingtao and Mou, Linzhan and Song, Linqi and Sun, Zhenan and Wei, Ying},
  journal={NIPS 2024},
  year={2024}
}

@article{chen2020survey,
  title={A survey of accelerator architectures for deep neural networks},
  author={Chen, Yiran and Xie, Yuan and Song, Linghao and Chen, Fan and Tang, Tianqi},
  journal={Engineering},
  volume={6},
  number={3},
  pages={264--274},
  year={2020},
  publisher={Elsevier}
}

@article{muralidharan2024compact,
  title={Compact language models via pruning and knowledge distillation},
  author={Muralidharan, Saurav and Turuvekere Sreenivas, Sharath and Joshi, Raviraj and Chochowski, Marcin and Patwary, Mostofa and Shoeybi, Mohammad and Catanzaro, Bryan and Kautz, Jan and Molchanov, Pavlo},
  journal={NIPS 2024},
  year={2024}
}

@article{benevs1964permutation,
  title={Permutation groups, complexes, and rearrangeable connecting networks},
  author={Bene{\v{s}}, V{\'a}clav E},
  journal={Bell System Technical Journal},
  volume={43},
  number={4},
  pages={1619--1640},
  year={1964},
  publisher={Wiley Online Library}
}

@inproceedings{takhar2022holl,
  title={HOLL: Program synthesis for higher order logic locking},
  author={Takhar, Gourav and Karri, Ramesh and Pilato, Christian and Roy, Subhajit},
  booktitle={TACAS 2022},
  year={2022}
}

@inproceedings{tang2019openfpga,
  title={OpenFPGA: An opensource framework enabling rapid prototyping of customizable FPGAs},
  author={Tang, Xifan and Giacomin, Edouard and Alacchi, Aur{\'e}lien and Chauviere, Baudouin and Gaillardon, Pierre-Emmanuel},
  booktitle={FPL 2019},
  year={2019}
}

\end{document}